\documentclass[aps,floatfix,groupedaddress,superscriptaddress,longbibliography,footinbib,notitlepage,showpacs,10pt]{revtex4-1}
\usepackage{fix-cm}
\usepackage{graphicx,graphics,times,bm,bbm,bbold,amssymb,soul}
\usepackage[T1]{fontenc}
\usepackage[utf8]{inputenc}
\usepackage{nicematrix,tikz}
 \usepackage{relsize}
\usepackage{yfonts}
\usepackage{mathrsfs}
\usepackage{enumitem}
\usepackage{amsthm,mathtools,amsmath,amsfonts,dsfont,color,xcolor,dcolumn}
\usepackage[bookmarksnumbered,colorlinks,plainpages]
{hyperref}
\hypersetup{
     colorlinks = true,
     citecolor  = blue,
     urlcolor   = blue}
     
\newcolumntype{P}[1]{>{\centering\arraybackslash}p{#1}}

\usepackage[normalem]{ulem}

\usepackage{listings}

\usepackage{tabularx,ragged2e}

\usepackage{soul}
\usepackage{ulem}
\usepackage{xcolor,cancel}

\begin{document}

\title{$\ell$-Multiranks of Multipartite Quantum States via Tensor Flattening: A Mathematica Codebase}

\author{Masoud Gharahi} 
\email[]{masoud.gharahi@gmail.com}
\email[]{masoud.gharahighahi@uj.edu.pl}
\affiliation{Faculty of Physics, Astronomy and Applied Computer Science, Jagiellonian University, 30-348 Kraków, Poland}


\begin{abstract}
We present a Mathematica codebase for computing $\ell$-multilinear ranks ($\ell$-multiranks) of multiqudit quantum states using tensor-flattening techniques. By calculating the ranks of all bipartition-induced matricizations, the method provides an efficient criterion for detecting Genuine Multipartite Entangled (GME) states in systems with local dimension $d$. The code automatically generates all required tensor reshapes and outputs the full $\ell$-multirank profile, offering a practical tool for characterizing entanglement in high-dimensional multiqudit systems.
\end{abstract}

\maketitle


\section{Introduction}

Quantum entanglement stands at the core of modern quantum information theory, serving as a fundamental resource for quantum communication, computation, and many-body physics \cite{HHHH09}. Although bipartite entanglement is now well understood, characterizing entanglement in multipartite systems remains considerably more challenging. As the number of subsystems and local dimensions increase, distinguishing between separable, biseparable, and genuinely multipartite entangled configurations becomes both computationally demanding and conceptually intricate.

One powerful approach to this problem arises from algebraic geometry and multilinear algebra, which provide structural tools for analyzing tensor representations of multipartite states. Quantum states of composite systems naturally correspond to high-order tensors, and properties such as separability and entanglement can be related to geometric features of the associated tensor spaces \cite{Harris, Landsberg}. Recent work has shown that $\ell$-multilinear ranks ($\ell$-multiranks)—obtained by flattening a tensor along different subsystem partitions—yield fine-grained invariants capable of distinguishing entanglement types beyond the bipartite regime \cite{GMO20, GM21, Gharahi24}. In particular, the $\ell$-multiranks capture the ranks of all matricizations defined by choosing $\ell$ subsystems versus their complement, and a state is genuinely multipartite entangled precisely when all such ranks are strictly greater than one.

Despite their conceptual utility, computing $\ell$-multiranks for arbitrary multiqudit states becomes tedious when performed manually, especially for high-dimensional Hilbert spaces or symbolic amplitudes. This motivates the development of computational tools capable of automating tensor flattening and rank extraction. In this paper, we introduce a Mathematica codebase \cite{Mathematica} designed to compute $\ell$-multiranks efficiently for general multiqudit systems. The package constructs all relevant bipartition-induced matricizations, computes their matrix ranks using standard linear-algebra routines \cite{Horn-Johnson}, and outputs the complete $\ell$-multirank profile of the given state. Our implementation offers a practical platform for exploring multipartite entanglement structures and supporting further developments at the intersection of quantum information theory and algebraic geometry.


\section{$\ell$-Multiranks}

In this work, we focus on pure quantum states of an $n$-partite composite system,
\begin{equation}\label{n-partite}
|\psi\rangle=\sum_{j=1}^{n}\sum_{i_j\in\mathbbm{Z}_{d_j}}
c_{i_1\cdots i_n}\,|i_1\cdots i_n\rangle ,
\end{equation}
which may be viewed as order-$n$ tensors living in the finite-dimensional Hilbert space
\begin{equation}
\mathcal{H}_{\delta}\coloneqq\bigotimes_{j=1}^{n}\mathbbm{C}^{d_j}\,,
\end{equation}
with $\delta\coloneqq d_1\cdots d_n$.
Here, the set of vectors 
\begin{equation}
\{|i_1\cdots i_n\rangle\equiv|i_1\rangle\otimes\cdots\otimes|i_n\rangle
\;\big|\; i_j\in\mathbbm{Z}_{d_j},\; 1\le j\le n\}
\end{equation}
form the standard computational basis of $\mathcal{H}_{\delta}$.

A pure state of the form \eqref{n-partite} is called \emph{fully separable} (or \emph{fully product}) if it factorizes into a tensor product of all local states, i.e., if there exist $|\varphi_j\rangle\in\mathbbm{C}^{d_j}$ for all $1\leq j \leq n$ such that
\begin{equation}
|\psi\rangle = |\varphi_1\rangle \otimes \cdots \otimes |\varphi_n\rangle .
\end{equation}
If no such factorization exists, the state is called \emph{entangled}.

An entangled $n$-partite pure state is said to be \emph{genuine multipartite entangled} (GME) if it does not decompose into a product state across any nontrivial bipartition. More precisely, for every bipartition $(S,\bar{S})$ with 
$S,\bar S\neq\varnothing$ and $S\cup\bar S=\{1,\dots,n\}$,
\begin{equation}
|\psi\rangle \neq |\phi\rangle_{S} \otimes |\chi\rangle_{\bar S}.
\end{equation}
Such states are also referred to as \emph{indecomposable}.

To assess whether a given multipartite state is GME, we invoke tools from multilinear algebra. Within this framework, the vectorization of an order-$n$ tensor, as presented in Eq.~\eqref{n-partite}, constitutes a particular instance of tensor reshaping. In the present work, we focus on a distinguished reshaping procedure—\emph{tensor flattening} (or \emph{matricization}) \cite{Landsberg}—which reorganizes the $n$-fold tensor product space $\mathcal{H}_{\delta}$ into a bipartite tensor product with enlarged local dimensions. To formalize the construction, let $I=(j_1,\ldots,j_{\ell})$ be an ordered $\ell$-tuple satisfying $1\leq \ell \leq \lfloor n/2 \rfloor$ and $1\leq j_1<\cdots<j_{\ell}\leq n$. Its ordered complement is denoted by $\bar{I}$, chosen such that $I\cup\bar{I}=\{1,2,\ldots,n\}$. This induces a canonical identification of the global Hilbert space with a bipartite decomposition
\begin{equation}
\mathcal{H}_{\delta}\simeq \mathcal{H}_{I}\otimes \mathcal{H}_{\bar{I}}\,,
\end{equation}
where $\mathcal{H}_{I}=\otimes_{\iota=j_1}^{j_{\ell}}\mathbbm{C}^{d_{\iota}}$ and $\mathcal{H}_{\bar{I}}$ denotes the complementary Hilbert space. For any pure vector $|\psi\rangle\in\mathcal{H}_{\delta}$, the $\ell$-partition $I$ naturally defines a linear map
\begin{equation}
\mathcal{M}_{I}[\psi]\colon \mathcal{H}_{I}^{\ast}\rightarrow \mathcal{H}_{\bar{I}},
\end{equation}
which we refer to as the matricization associated with $I$. Expressed in Dirac notation, this map admits the representation
\begin{equation}
\mathcal{M}_{I}[\psi]
=\bigl(\langle e_{1}|\psi\rangle,\ldots,\langle e_{d_I}|\psi\rangle\bigr)^{\mathrm{T}},
\end{equation}
where $\{|e_m\rangle = |i_{j_1}\cdots i_{j_\ell}\rangle \mid i_{j_k}\in\mathbbm{Z}_{d_{j_k}},~1\leq k\leq\ell\}_{m=1}^{d_I}$ is the canonical basis of $\mathcal{H}_{I}$, and $d_I=\prod_{\iota=j_1}^{j_{\ell}} d_{\iota}$. The symbol ${\mathrm{T}}$ denotes matrix transposition.

Since each admissible $\ell$-tuple $I$ induces a distinct matricization, a state $|\psi\rangle$ gives rise to ${\binom{n}{\ell}}$ such matrix representations. This observation motivates the definition of the \emph{$\ell$-multiranks} of $|\psi\rangle$, namely the ${\binom{n}{\ell}}$-tuple consisting of the ranks of all matrices $\mathcal{M}_{I}[\psi]$ \cite{GMO20,GM21}.


\section{Mathematica Codebase}

In this section, we present the \texttt{Mathematica} routines used to compute $\ell$-multiranks of multiqudit quantum states via tensor flattening. Two implementations are provided: one based on \texttt{SparseArray} representations of the state tensor, and another using explicit \texttt{TensorProduct} constructions. Both approaches automatically generate all bipartition-induced matricizations and return the associated matrix ranks. The full source code is publicly available online \cite{mathoud-github}.

\subsection{Flattening (Code~1: SparseArray)}

\begin{lstlisting}[language=Mathematica,caption={Flattening using \texttt{SparseArray}}]
n = #; (* # indicates the number of parties. *)

subb = Table[Subsets[Range[n], {i}], {i, 1, Floor[n/2]}];
T = SparseArray[{a}, Table[2, n]];
(* instead of a, write the nonzero elements of your state. *)
(* 2 is the qubit dimension. Change it to your desired d for higher dimensions. *)

Flatten[
  Table[
    Module[{partition = Part[subb, i, j], rest},
      rest = Complement[Range[n], partition];
        MatrixRank[Flatten[T, {partition, rest}]] ],
          {i, Floor[n/2]}, {j, Binomial[n, i]} ], {1} ]
\end{lstlisting}

\subsection*{Examples Using \texttt{SparseArray}}

\subsubsection*{Three Qubits}

In this example, we consider the tensor $T$ corresponding to the three-qubit state $|001\rangle + |010\rangle + |100\rangle$.

\begin{lstlisting}[language=Mathematica]
n = 3;
subb = Table[Subsets[Range[n], {i}], {i, 1, Floor[n/2]}];

T = SparseArray[{{1, 1, 2} -> 1, {1, 2, 1} -> 1, {2, 1, 1} -> 1},Table[2, n]];

Flatten[
  Table[
    Module[{partition = Part[subb, i, j], rest},
      rest = Complement[Range[n], partition];
        MatrixRank[Flatten[T, {partition, rest}]] ],
          {i, Floor[n/2]}, {j, Binomial[n, i]} ], {1} ]
        
Output: {{2, 2, 2}}
\end{lstlisting}

\subsubsection*{Four Qubits}

Here, $T$ represents the four-qubit state
$|0000\rangle + |0011\rangle + |1100\rangle - |1111\rangle$.

\begin{lstlisting}[language=Mathematica]
n = 4;
subb = Table[Subsets[Range[n], {i}], {i, 1, Floor[n/2]}];

T = SparseArray[{{1, 1, 1, 1} -> 1,{1, 1, 2, 2} -> 1,{2, 2, 1, 1} -> 1,{2, 2, 2, 2} -> -1},Table[2, n]];

Flatten[
  Table[
    Module[{partition = Part[subb, i, j], rest},
      rest = Complement[Range[n], partition];
        MatrixRank[Flatten[T, {partition, rest}]] ],
          {i, Floor[n/2]}, {j, Binomial[n, i]} ], {1} ]
        
Output: {{2, 2, 2, 2}, {2, 4, 4, 4, 4, 2}}
\end{lstlisting}

\subsection{Flattening (Code~2: TensorProduct)}

\begin{lstlisting}[language=Mathematica,caption={Flattening using \texttt{TensorProduct}}]
n = #; (* # indicates the number of parties. *)

subb = Table[Subsets[Range[n], {i}], {i, 1, Floor[n/2]}];

(* Define the basis. Example (standard qubit basis): p0 = {1, 0}; p1 = {0, 1}; *)

(* Write your state using TensorProduct. *)

Flatten[
  Table[
    Module[{partition = Part[subb, i, j], rest},
      rest = Complement[Range[n], partition];
        MatrixRank[Flatten[T, {partition, rest}]] ],
          {i, Floor[n/2]}, {j, Binomial[n, i]} ], {1} ]
\end{lstlisting}

\subsection*{Examples Using \texttt{TensorProduct}}

\subsubsection*{Three Qutrits}

In this example, we consider the tensor $T$ as a three-qutrit state
$|002\rangle + |020\rangle + |200\rangle + |011\rangle + |101\rangle + |110\rangle$.

\begin{lstlisting}[language=Mathematica]
n = 3;
subb = Table[Subsets[Range[n], {i}], {i, 1, Floor[n/2]}];

p0 = {1, 0, 0};
p1 = {0, 1, 0};
p2 = {0, 0, 1};

T = TensorProduct[p0, p0, p2] +
    TensorProduct[p0, p2, p0] +
    TensorProduct[p2, p0, p0] +
    TensorProduct[p0, p1, p1] +
    TensorProduct[p1, p0, p1] +
    TensorProduct[p1, p1, p0];

Flatten[
  Table[
    Module[{partition = Part[subb, i, j], rest},
      rest = Complement[Range[n], partition];
        MatrixRank[Flatten[T, {partition, rest}]] ],
          {i, Floor[n/2]}, {j, Binomial[n, i]} ], {1} ]
        
Output: {{3, 3, 3}}
\end{lstlisting}

\subsubsection*{Six Qutrits}

In this example, the tensor $T$ corresponds to the six-qutrit state
$|0\rangle^{\otimes 6} + |1\rangle^{\otimes 6} + |2\rangle^{\otimes 6} + |001122\rangle$.

\begin{lstlisting}[language=Mathematica]
n = 6;
subb = Table[Subsets[Range[n], {i}], {i, 1, Floor[n/2]}];

p0 = {1, 0, 0};
p1 = {0, 1, 0};
p2 = {0, 0, 1};

T = TensorProduct @@ ConstantArray[p0, n] +
    TensorProduct @@ ConstantArray[p1, n] +
    TensorProduct @@ ConstantArray[p2, n] +
    TensorProduct[p0, p0, p1, p1, p2, p2];

Flatten[
  Table[
    Module[{partition = Part[subb, i, j], rest},
      rest = Complement[Range[n], partition];
        MatrixRank[Flatten[T, {partition, rest}]] ],
          {i, Floor[n/2]}, {j, Binomial[n, i]} ], {1} ]

Output: {{3, 3, 3, 3, 3, 3},{3, 4, 4, 4, 4, 4, 4, 4, 4, 3, 4, 4, 4, 4, 3},
         {4, 4, 4, 4, 4, 4, 4, 4, 4, 4, 4, 4, 4, 4, 4,4, 4, 4, 4, 4}}
\end{lstlisting}


\section*{Acknowledgments}
The author acknowledges financial support from the European Union under ``ERC Advanced Grant TAtypic, Project No.~101142236''. The views and opinions expressed are, however, those of the author only and do not necessarily reflect those of the European Union or the European Research Council Executive Agency. Neither the European Union nor the granting authority can be held responsible for them.


\end{document}